\documentclass[preprint,aps]{revtex4}

\usepackage{graphicx}
\usepackage{dcolumn}
\usepackage{bm}
\usepackage{epsf}

\newcommand{\be}{\begin{eqnarray}}

\newcommand{\Eq}[1]{Eq.~(\ref{#1})}

\newcommand{\ur}[1]{(\ref{#1})}

\newcommand{\beq}{\begin{equation}}
\newcommand{\eeq}{\end{equation}}

\newcommand{\la}[1]{\label{#1}}
\newcommand{\bea}{\begin{eqnarray}}
\newcommand{\eea}{\end{eqnarray}}
\newcommand{\beqa}{\begin{eqnarray}}
\newcommand{\eeqa}{\end{eqnarray}}
\newcommand{\ba}{\begin{array}}
\newcommand{\ea}{\end{array}}

\newcommand{\half}{{\textstyle{\frac{1}{2}}}}

\newcommand{\Th}{$\Theta^+\,$}

\def\appendix{\par
\setcounter{subsection}{0}
\setcounter{equation}{0}

\def\thesection{Appendix}
\def\theequation{\Alph{section}.\arabic{equation}}}

\begin{document}

\title{\bf Baryon resonances in the mean field approach \\
and \\
a simple explanation of the $\Theta^+$ pentaquark}

\author{
\bf Dmitri Diakonov$^{1,2}$
}

\affiliation{
$^1$Petersburg Nuclear Physics Institute, Russia \\
$^2$Institut f\"ur Theoretische Physik II, Ruhr-Universit\"at, Bochum, Germany
}

\date{December 17, 2008}

\begin{abstract}
We suggest to classify baryon resonances as single-quark states in a mean field,
and/or as its collective excitations. Identifying the Roper resonance $N(1440,\half^+)$,
the nucleon resonance $N(1535,\half^-)$, and the singlet hyperon $\Lambda(1405,\half^-)$
as single-quark excitations, we find that there must be an exotic $S=+1$ baryon
resonance \Th (the ``pentaquark'') with a mass about $1440+1535-1405=1570$ MeV
and spin-parity $\half^+$. We argue that \Th is an analog of the Gamov--Teller
excitation long known in nuclear physics.

\end{abstract}

\maketitle

It was argued 30 years ago by Witten~\cite{Witten-Nc} that if the number of colors $N_c$ is large,
the $N_c$ quarks constituting a baryon can be viewed as moving in a mean field whose fluctuations
are suppressed as $1/N_c$. Whether $N_c\!=\!3$ of the real world is large enough for the mean
field in baryons to be a working notion is a question to which there is no general answer:
it depends on how large are $1/N_c,\,1/N_c^2,...$ corrections to a particular baryon observable.
However, experience in hadron physics tells us that usually the relations between observables found
in the large-$\!N_c$ limit are in satisfactory agreement with reality, unless there are special
reasons to expect large $1/N_c$ corrections~\cite{Witten-Nc,Manohar}. In any case, it is helpful
to understand how baryons are constructed in the large-$\!N_c$ limit, before corrections are
considered.

The mean field can, in principle, have components with various quantum numbers
$J^{PC}=0^{++}$ (scalar), $0^{-+}$ (pseudoscalar), $1^{--}$ (vector), $1^{++}$ (axial), etc.
Since the mean field is created by $N_c$ quarks of the $u,d,s$ flavor it is also characterized
by the flavor quantum numbers like isospin $T$ and hypercharge $Y$. We do not consider baryons
with heavy quarks here.

One expects that the mean field inside heavy ground-state baryons has the maximal possible,
that is spherical symmetry. There is no problem in writing a spherical-symmetric scalar,
flavor-singlet field as $\;\sigma({\bf x})=P_1(r)$ where $r=|{\bf x}|$ is a distance
from the center of a baryon, however we immediately run into a problem of how to write
the {\it Ansatz} for, say, the mean pseudoscalar field. Being pseudoscalar it has to be
odd in ${\bf x}$. The minimal extension of spherical symmetry is then the ``hedgehog''
{\it Ansatz} ``marrying'' the isotopic and space axes:
\beq
\pi^a({\bf x})=n^a\,P_2(r),\quad n^a=\frac{x^a}{r},\quad a=1,2,3;\qquad \pi^{4,5,6,7,8}({\bf x})=0.
\la{hedgehog}\eeq
This {\it Ansatz} breaks spontaneously the symmetry under independent space and isospin rotations,
and only a simultaneous rotation in both spaces remains a symmetry. At the same time it
breaks the $SU(3)$ flavor symmetry. One may argue that the $SU(3)$ symmetry is explicitly
broken from the start by $m_s\gg m_{u,d}$, however one can as well consider the strange
quark mass $m_s$ as a small perturbation~\cite{footnote-1}. In the chiral limit,
$m_s\to 0$, the {\it Ansatz} \ur{hedgehog} breaks spontaneously
the $SU(3)$ symmetry: the first three component of the pseudoscalar octet are privileged. 
Full symmetry is restored when one rotates the asymmetric mean field in flavor 
and ordinary spaces: that produces many baryon states with definite quantum numbers.

The r\^ole of a small $m_s$ is in fact the same as of the infinitesimal magnetic field
in materials with a spontaneous magnetization: it establishes the preferred direction of
magnetization. In our case, the {\it Ansatz} \ur{hedgehog} is privileged since
$m_s\neq 0$, regardless of whether it is considered sizable or infinitesimal.

If in a baryon there are mean vector fields with the quantum numbers of $\omega$ and $\phi$
(and there are no {\it a priori} reasons why they should be absent), the {\it Ans\"atze} for
those fields in correspondence with \Eq{hedgehog} are $\;\omega_0,\phi_0({\bf x})=P_{3,4}(r);\;
\omega_i,\phi_i({\bf x})=n_i\,P_{5,6}(r)$. The {\it Ansatz} for the axial $a_1$ field is
$\;A^a_0=n^aP_7(r),\;A^a_i=\epsilon_{aij}n_jP_8(r)+\delta^a_iP_9(r)+n^an_iP_{10}(r)$,
and so on.

In the mean field approximation, justified at large $N_c$, one looks for the solutions
of the Dirac equation for single quark states in the background mean field.
The Dirac Hamiltonian for quarks is, schematically,
\beq
H=\gamma^0\left(i\gamma^i\partial_i+\sigma({\bf x})+i\gamma^5\pi({\bf x})
+\gamma^\mu V_\mu({\bf x})+\gamma^\mu\gamma^5A_\mu({\bf x})+\ldots\right)
\la{DiracH}\eeq
In fact, the one-particle Dirac Hamiltonian is most probably nonlocal and momentum-dependent (as
it would follow {\it e.g.} from Fiertz-transforming and then bosonizing color quark interactions),
therefore \Eq{DiracH} is a symbolic presentation. However, several important statements can be made
on general grounds:
\begin{itemize}
\item Given the {\it Ansatz} for the mean fields $\sigma,\pi,V,A$, the
Hamiltonian \ur{DiracH} actually splits into two: one for $s$ quarks ($H_s$) and the other
for $u,d$ quarks ($H_{ud}$). The former commutes with the angular momentum of $s$ quarks,
${\bf J}={\bf L}+{\bf S}$, and with the inversion of spatial axes, hence all energy levels
of $s$ quarks are characterized by half-integer $J^P$ and are $(2J+1)$-fold degenerate.
The latter commutes only with the `grand spin' ${\bf K}={\bf T}+{\bf J}$ and with inversion,
hence the $u,d$ quark levels have definite integer $K^P$ and are $(2K+1)$-fold degenerate.
The energy levels for $u,d$ quarks on the one hand and for $s$ quarks on the other are
completely different, even in the chiral limit $m_s\to 0$
\item The Hamiltonian \ur{DiracH} has mixed symmetry with respect to time reversal, therefore
the one-particle spectra for $s$ and $u,d$ quarks are generally not symmetric under
the change $E\to -E$
\item All energy levels, both positive and negative, are probably discrete owing to confinement.
Indeed, a continuous spectrum would correspond to a situation when quarks are free at large
distances from the center, which contradicts confinement. [One can mimic or model confinement
for example by imposing the condition that the effective quark mass $\sigma({\bf x})$ grows at infinity.]
\item According to the Dirac theory, all {\em negative}-energy levels, both for $s$ and $u,d$ quarks,
have to be fully occupied, corresponding to the vacuum. It means that there must be exactly
$N_c$ quarks antisymmetric in color occupying all (degenerate) levels with $J_3$ from $-J$ to $J$,
or $K_3$ from $-K$ to $K$; they form closed shells that do not carry quantum numbers
\item Filling in the lowest level with $E>0$ by $N_c$ quarks makes a baryon.
\end{itemize}

We shall suppose that the energy levels with minimal $|E|$ are $K^P=0^+$ for $u,d$ quarks
and $J^P=\half^+$ for $s$ quarks, because it corresponds to the maximal-symmetry wave functions
as it is usual for the Dirac equation in smooth external fields. {\it A priori} the signs of $E$
for those solutions can be any, however experimentally the lowest baryon is a nucleon
and not the $\Omega^-$ hyperon. It means, that the $\half^+$ level for $s$ quarks
is probably the nearest to $E=0$ but remains on the negative side. It belongs to the
vacuum sector and has to be filled in together with all the rest negative-energy levels.
On the contrary, the $0^+$ level for $u,d$ quarks must be the lowest at $E>0$.
Filling it in with $N_c$ quarks antisymmetric in color, adds $N_c$ $\;u,d$ quarks to the vacuum,
and makes the nucleon, see Fig.~1. We do not know yet how much higher is the highest filled
$u,d$ shell than the highest filled $s$-quark shell but shall determine in shortly from 
experiment (at $N_c=3$).

\begin{figure}[htb]
\begin{minipage}[t]{.45\textwidth}
\includegraphics[width=\textwidth]{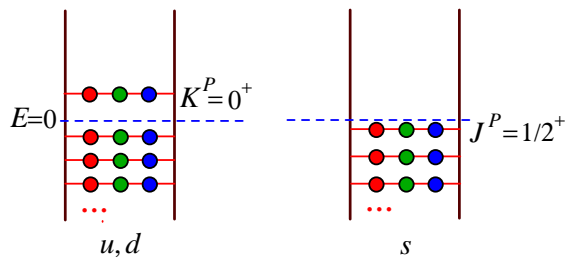}
\end{minipage}
\caption{Filled quark levels for the ground-state baryon $N(940,\half^+)$.
The two lightest baryon multiplets $({\bf 8},\half^+)$ and $({\bf 10},\frac{3}{2}^+)$
are rotational excitations of the same filling scheme.}
\label{fig:1}

\end{figure}

The true quantum numbers of the lightest baryons are determined from the quantization
of the rotations of the mean field since the {\it Ansatz} discussed above
spontaneously breaks symmetry under rotations in ordinary and flavor spaces. However, a simultaneous
rotation in ordinary and isospin spaces remains a symmetry. Therefore, if one limits oneself
to non-strange baryons, the quantization of rotations produces states with $J\!=\!T\!=\!\half$
(for any odd $N_c$), that is the nucleon, and $J\!=\!T\!=\!\frac{3}{2}$, the $\Delta$ resonance~\cite{Witten-Sk}. 
If $m_s$ is treated as a perturbation~\cite{footnote-1}, one has to extend this to $SU(3)$ flavor rotation. Its quantization
gives specific $SU(3)$ multiplets that reduce at $N_c\!=\!3$ to the octet with spin $\half$ and the decuplet with
spin $\frac{3}{2}\;$, see {\it e.g.}~\cite{DP-08}. Witten's quantization condition $Y'\!=\!\frac{N_c}{3}$~\cite{Witten-Sk}
follows trivially from the fact that there are $N_c$ $\;u,d$ valence quarks each with the hypercharge
$\frac{1}{3}$~\cite{Blotz}. Therefore, the ground state shown in Fig.~1 entails in fact 56 rotational
states. It is the same ${\bf 56}$-plet as the ground shell of the nonrelativistic quark model
but its interpretation is different. The splitting between the centers of the multiplets $({\bf 8},\half^+)$
and $({\bf 10}, \frac{3}{2}^+)$ is ${\cal O}(1/N_c)$, and the splittings inside multiplets
can be to determined as a linear perturbation in $m_s$~\cite{Blotz}.

The picture has a similarity in nuclear physics where at large atomic numbers $A$, $\;Z$ protons
and $A\!-\!Z$ neutrons are considered in different self-consistent mean fields and have a
different system of one-particle levels. One fills proton and neutron levels separately up
to the common Fermi surface. Contrary to the quark case, the negative-energy levels for nucleons can be
neglected because of the large nucleon mass. There are collective excitations of heavy nuclei,
{\it e.g.} rotation whose energy scale is ${\cal O}(1/A)$ (in the baryon case it scales as $1/N_c$).
However, there are also one-particle and particle-hole excitations that are of the order of unity
in $A$. Similarly, one should expect ${\cal O}(N_c^0)$, that is large, one-particle and particle-hole
excitations. Let us try to identify them.

The lowest baryon resonance beyond the rotational excitations of the ground state
is the singlet $\Lambda(1405,\half^-)$. Apparently, it can be obtained only as an excitation
of the $s$ quark, and its quantum numbers must be $J^P=\half^-$.
The resonance $\Lambda(1405,\half^-)$ is excited if one of the $u,d$ quarks from the valence
$0^+$ level jumps, under the action of a $S=-1$ force, to the first excited state for $s$ quarks,
see Fig.~2. It is predominantly a 3-quark state (at $N_c\!=\!3$). This excited state
generates a rotational band of $SU(3)$ multiplets of its own, but we do not consider them
here.
\begin{figure}[htb]
\begin{minipage}[t]{.48\textwidth}
\includegraphics[width=\textwidth]{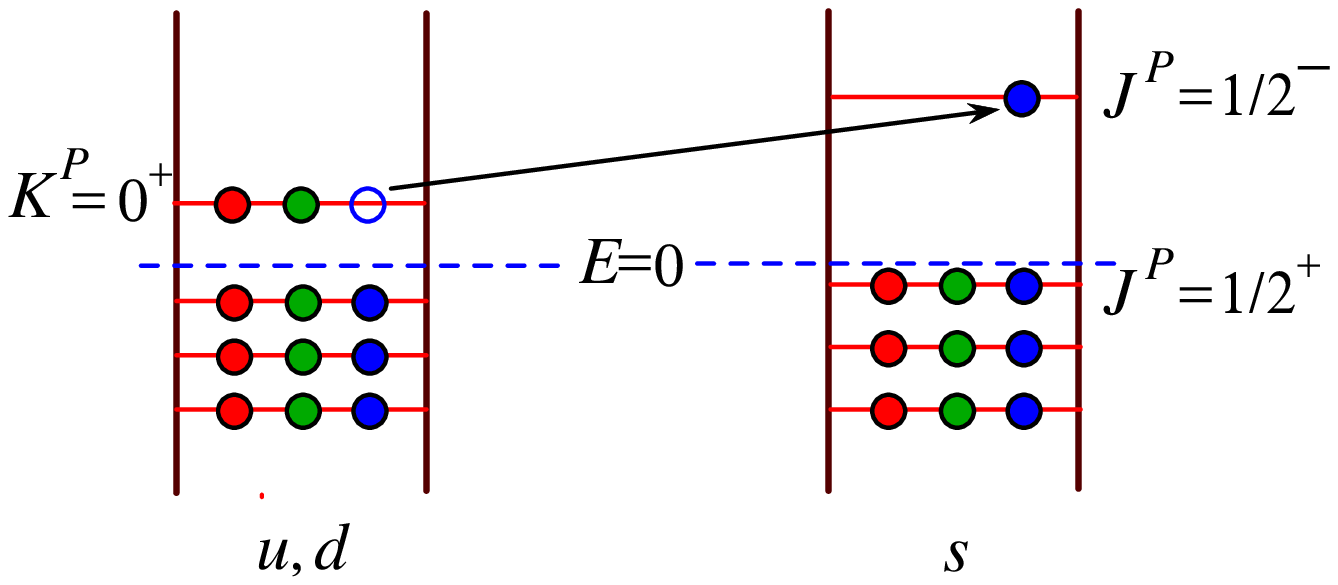}
\caption{$\Lambda(1405,\half^-)$}
\label{fig:2}
\end{minipage}
\hfil
\begin{minipage}[t]{.48\textwidth}
\includegraphics[width=\textwidth]{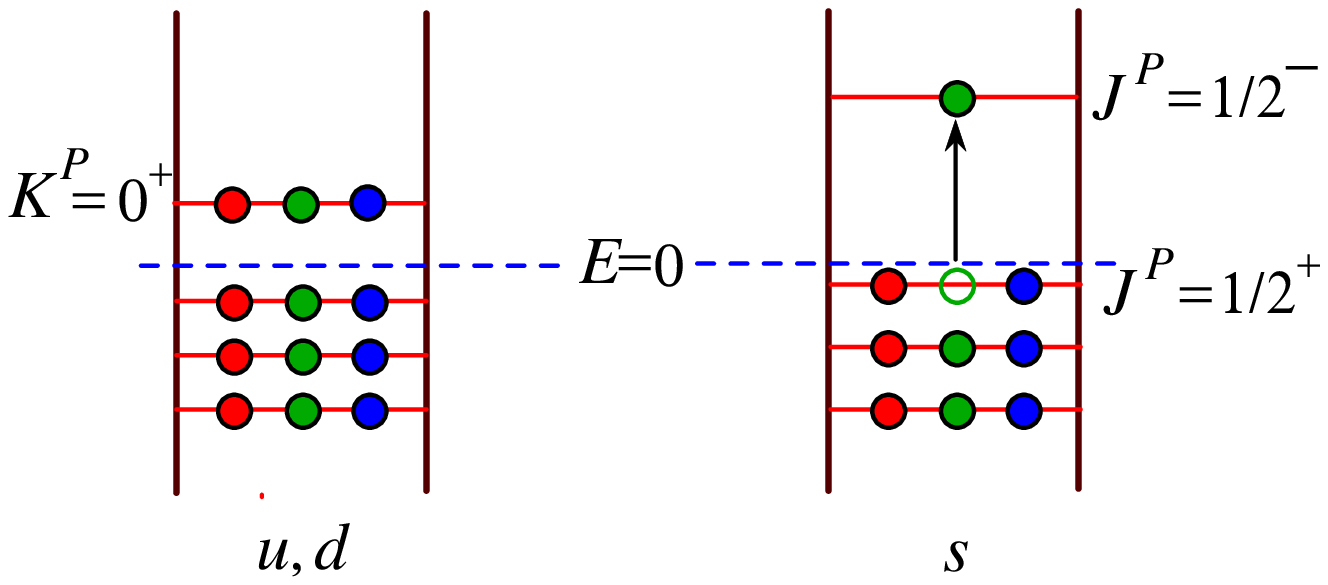}
\caption{$N(1535,\half^-)$}
\label{fig:3}
\end{minipage}
\end{figure}

If there is a $\half^-$ level for $s$ quarks, it can be excited also by an
$s$ quark jumping from its highest filled level $\half^+$, see Fig.~3. This is a particle-hole
excitation which does not change the nucleon quantum numbers (except for parity), as we
are just adding an $s\bar s$ pair to the valence $u,d$ level that determines the quantum numbers.
We therefore identify this excitation with $N(1535,\half^-)$. We see that at $N_c\!=\!3$
it is predominantly a pentaquark state $u(d)uds\bar s$. That explains its large branching ratio
in the $\eta N$ decay~\cite{Zou}, a long-time mystery. We also see that, since the highest filled level for
$s$ quarks is lower than the highest filled level for $u,d$ quarks, $N(1535,\half^-)$
must be {\em heavier} than $\Lambda(1405\half^-)$: the opposite prediction of the nonrelativistic
quark model has been always of some concern. We stress that in our picture the existence
of an unusual nucleon resonance $N(1535,\half^-)$ is a {\em consequence} of the $\Lambda(1405,\half^-)$
existence. The transition shown in Fig.~3 also entails its own rotational band.
Subtracting $1535-1405=130$, we find that the $\half^+$ $s$-quark level is approximately
130 MeV lower in energy than the valence $0^+$ level for $u,d$ quarks.

There is also a low-lying Roper resonance $N(1440,\half^+)$. It requires that there is
an excited one-particle $u,d$ state with $K^P=0^+$, see Fig.~4. Just as the
ground state nucleon, it is part of the excited $({\bf 8'},\half^+)$
and $({\bf 10'}, \frac{3}{2}^+)$ split as $1/N_c$. In fact the first excited state could
be also $K^P=1^+,2^+$ which would generate more $SU(3)$ multiplets including one
with the Roper resonance; $K^P=0^+$ is a minimal hypothesis. The identification
of the nature of the Roper resonance solves another problem of the nonrelativistic model
where $N(1440,\half^+)$ must be heavier than $N(1535,\half^-)$. In our approach they
are simply unrelated. 

\begin{figure}[htb]
\begin{minipage}[t]{.45\textwidth}
\includegraphics[width=\textwidth]{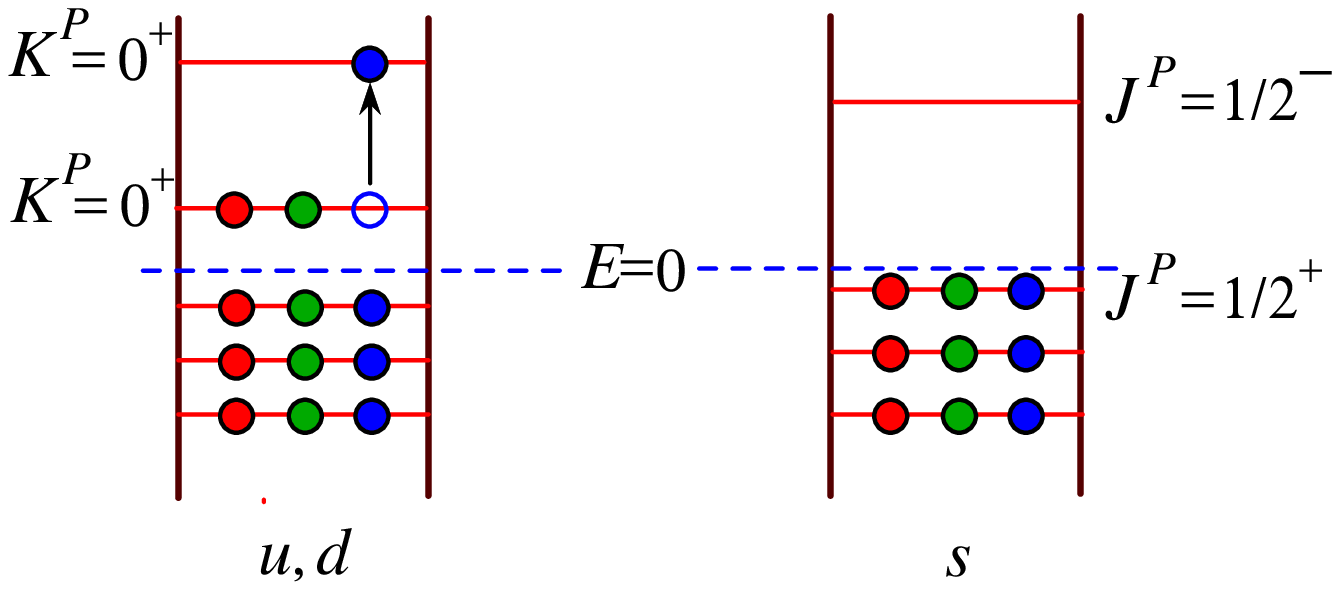}
\caption{$N(1440,\half^+)$}
\label{fig:4}
\end{minipage}
\hfil
\begin{minipage}[t]{.45\textwidth}
\includegraphics[width=\textwidth]{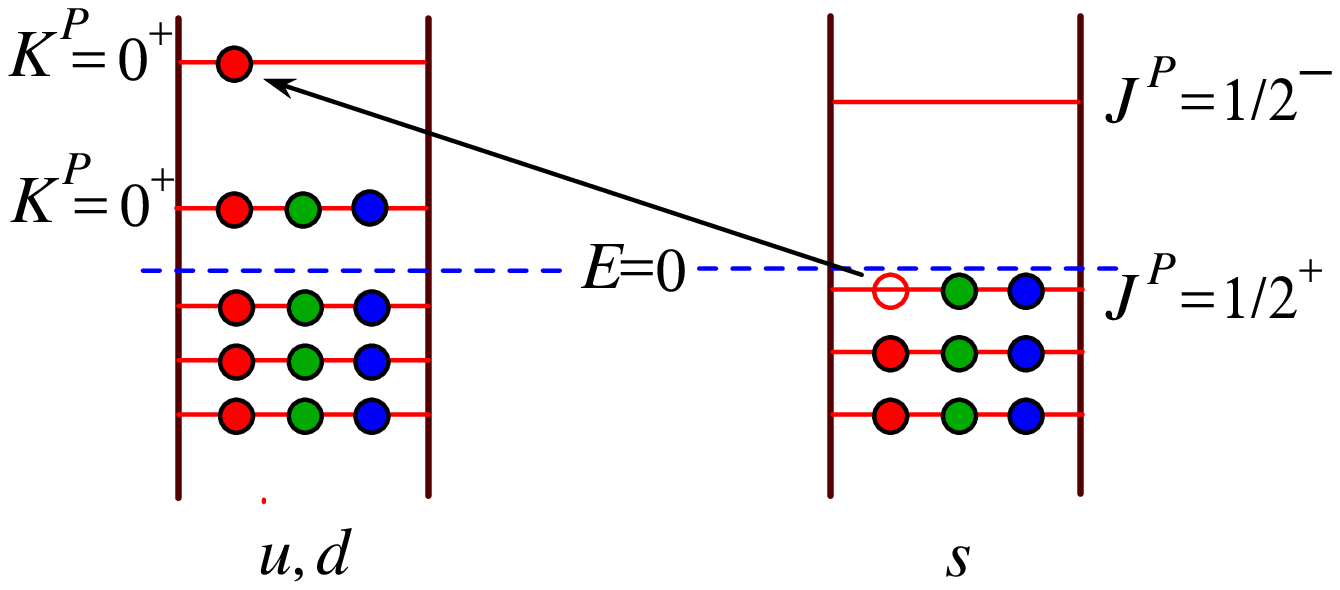}
\caption{$\Theta^+(\half^+)$}
\label{fig:5}
\end{minipage}
\end{figure}

We now come to the crucial point: Given that there is an unoccupied level for
$u,d$ quarks, one can put there an $s$ quark as well, taking it from one of the filled $s$-quark shells.
The minimal-energy excitation is from the highest occupied shell for $s$ quarks, to
the lowest unoccupied level for $u,d$ quark, that is to the would-be Roper level, see
Fig.~5. It is a particle-hole excitation with the valence level left untouched, its
quantum numbers being apparently $S=+1,\;T=0,\;J^P=\half^+$. At $N_c\!=\!3$ this excitation
is a pentaquark state $uudd\bar s$, precisely the exotic $\Theta^+$ baryon predicted in Ref.~\cite{DPP-97}
from other considerations. The quantization of its rotations in flavor and ordinary spaces produces
the antidecuplet $({\bf \overline{10}},\half^+)$ and higher multiplets
$({\bf 27},\frac{3}{2}^+)$ and $({\bf 27},\frac{1}{2}^+)$~\cite{DP-04}.

Since the relative position of all four levels involved are already known from the masses
of the well-established resonances $N(1440,\half^+),\;N(1535,\half^-)$ and $\Lambda(1405,\half^-)$,
it is a matter of trivial arithmetics to find the energy difference between the $s$-quark shell and
the first excited $u,d$-quark level. We obtain an estimate for the \Th mass:
$m_{\Theta}\approx 1440+1535-1405=1570\,{\rm MeV}$. Of course, one should not understand
the number literally. First, the masses of the resonances exploited here
are known with an uncertainty of a few tens MeV. Second, the masses of physical resonances
are not fixed exactly by the one-particle levels but get ${\cal O}(1/N_c)$ and ${\cal O}(m_s)$
corrections (which in principle are calculable). Nevertheless, the most interesting prediction that
the exotic pentaquark is a consequence of the three well-known resonances
and must be light, is an unambiguous feature of the picture.

In nuclear physics, the charge-exchange excitations generated by the axial current $j_{\mu\,5}^\pm$,
when a neutron from the last occupied shell is sent to an unoccupied proton level or {\it vice versa}
are known as Gamov--Teller transitions~\cite{BM}; they have been extensively studied both
experimentally and theoretically. Thus our interpretation of the \Th is that it is a Gamov--Teller-type
resonance long known in nuclear physics.

However, the calculation of the \Th width should be different from that in nuclear physics.
For the Gamov--Teller transition in a nuclei, it is sufficient to calculate the
matrix element between the initial and final states of the transition axial current. To make
it more accurate, one can take into account a correction from the admixture of 
particle-hole states to the one-particle states. This approximation is often called the RPA;
it goes beyond the mean field. 

In the baryon case, even a one-particle state of the leading mean-field
approximation, shown in Fig.~1, is in fact a Fock state with additional quark-antiquark pairs. 
These arise when one decomposes the filled negative-energy levels in the plane wave basis~\cite{PP-03,DP-05}.
Therefore, unlike its nuclear physics counterpart, the \Th decay amplitude has, already
in the mean field approximation, two contributions: one is from the conventional
``fall-apart'' process whereas the other is the ``5-to-5'' transition of the \Th to the 5-quark
component of the nucleon~\cite{D-06,DP-08}. The two amplitudes are separately not Lorentz
invariant -- only their sum is. In the lab frame there is a tendency for the two amplitudes
to cancel each other~\cite{Hosaka}. In the infinite momentum frame, however, the ``fall-apart''
amplitude (the simple one) is zero in the chiral limit, and only the ``5-to-5'' amplitude
survives. The one-particle Hamiltonian \ur{DiracH} is covariant, such that there is no
problem in transforming the mean field to an infinite momentum frame, which is the shortest
way to evaluate the \Th width. [It may seem somewhat unusual but we do not have much
experience from the past in computing pentaquarks widths!] The program has been carried out
in the Chiral Quark Soliton Model with the result $\Gamma_\Theta \sim 1\,{\rm MeV}$
with no parameter fitting~\cite{DP-05,ThetaWidth}.

It should be stressed that the small width of \Th has no relation to and no influence
from the large width of the Roper resonance. When \Th decays, the valence level (to which the
Roper resonance decays) is Pauli-blocked, and the fact that an ordinary ``3-to-3'' decay
width of the Roper resonance is large does not effect the width of the \Th which is
very narrow owing to physics completely different from that governing the Roper decay.

There can be additional one-particle and/or particle-hole excitations, however just the two
excited levels suggested here ($0^+$ for $u,d$ quarks and $\half^-$ for $s$ quarks) are sufficient
to explain the majority of baryon resonances up to 2 GeV, if the rotational states generated
by each of the excitations are taken into account. A detailed study of the ensuing baryon spectrum
will be published separately. For high-spin resonances (actually for $J\geq N_c$)
it may become energetically favorable to depart from a spherically-symmetric mean field.\\

I am grateful to L.~Frankfurt, K.~Goeke, A.~Hosaka, M.~Polyakov and especially to V.~Petrov for helpful
discussions. I thank German Research Society (DFG) for Mercator Professorship at Bochum.
This work has been partly supported by Russian Government grants RFBR-06-02-16786 and RSGSS-3628.2008.2.


\begin{thebibliography}{99}
\bibitem{footnote-1}
The lowest $SU(3)$ baryon multiplets satisfy Gell-Mann--Okubo relations with a high accuracy,
meaning that the splittings inside the multiplets are linear in $m_s$, actually
${\cal O}(m_sN_c)$. The splitting between the centers of the octet and of the decuplet
is ${\cal O}(\Lambda / N_c)$ where $\Lambda$ is the QCD scale parameter.
Numerically, the former quantity is about 140 MeV whereas the latter is $1382-1152=230$ MeV,
slightly larger. It means that in baryon physics one can indeed treat $m_s$ as a
very small quantity: $m_s={\cal O}(\Lambda/N_c^2)$ or even less.
\end{thebibliography}
\end{document}